\documentclass[12pt]{article}\usepackage{graphicx,amsmath,amssymb,cite}
\usepackage{epsfig,epsf}
\usepackage{graphicx}
\usepackage{epstopdf}
\pdfoutput=1

\newcommand{\beq}{\begin{equation}}
\newcommand{\eeq}{\end{equation}}
\newcommand{\alphabar}{\bar{\alpha}_s}
\newcommand{\betabar}{\bar{\beta_0}}

\numberwithin{equation}{section}

\begin{document}

\vspace*{0.5 cm}

\begin{center}

{\Large{\bf The Effect of a Rapidity Gap Veto on the Discrete BFKL Pomeron}}

\vspace*{1 cm}

{\large Douglas~A.~Ross~$^1$ and Agust{\'\i}n~Sabio~Vera~$^2$} \\ [0.5cm]
{\it $^1$ School of Physics \& Astronomy, University of Southampton,\\Highfield, Southampton SO17 1BJ, UK}\\[0.1cm]
{\it $^2$ Instituto de F{\' \i}sica Te\'{o}rica UAM/CSIC, Nicol{\' a}s Cabrera 15\\ \& Universidad Aut{\' o}noma de Madrid, E-28049 Madrid, ~Spain}\\[0.1cm]
 \end{center}

\vspace*{3 cm}

\begin{center}
{\bf Abstract} 
\end{center}  
We investigate the sensitivity of the discrete BFKL spectrum, which appears in the gluon Green function when the running coupling is considered, to a lower cut-off in the relative rapidities of the emitted particles.  
We find that the eigenvalues associated to each of the discrete eigenfunctions decrease with the size of the rapidity veto. The effect is stronger on the lowest eigenfunctions. The net result is a reduction of the growth with energy for the Green function together with a suppression in the regions with small transverse momentum.


\vspace*{3 cm}

\begin{flushleft}
  May 2016 \\ 
\end{flushleft}

\newpage

\section{Introduction}
In perturbative QCD, the counterpart of the pomeron of Regge theory
 is described in terms of a Green function ${\cal G}(Y,t,t^\prime)$
describing the rapidity, $Y$, dependence of the scattering amplitude 
 of a gluon with transverse momentum $k_T=\Lambda_{QCD} e^{t/2}$
and a gluon with transverse momentum $k^\prime_T=\Lambda_{QCD} e^{t^\prime/2}$
with a relative rapidity difference $Y$ between the two gluons. It is obtained \cite{BFKL} by   resumming the leading rapidity contributions 
to all orders in perturbation theory. At leading order, this is obtained by assuming a cascade of gluons emitted between the two
 primary gluons in the kinematic regime  in which the emitted gluons have a large rapidity relative to the preceding emitted
gluons.  Schmidt \cite{Schmidt} pointed out that a significant reduction in the resultant Green function occurs if one imposes this
 restriction explicitly by demanding that one only considers contributions to the scattering amplitude  in which emitted gluons 
 have a minimum rapidity gap, $b$, relative to the preceding emitted gluon. It was furthermore shown in ref.\cite{FRS} that
 the large effect of imposing such a restriction simulates, to a good approximation, the effect of the NLO corrections to the
 BFKL Green-function with collinear summation as proposed by Salam \cite{salam}. In particular the optimal match was found
 if one takes the resummation scheme 4 of \cite{salam} and  a rapidity gap veto (minimum rapidity gap between adjacent emitted
 gluons) $b \approx 2$. This is consistent with the original presentation of this idea by Lipatov in~\cite{LevTalk}. A rapidity veto has been used in different works also for non-linear evolution equations~\cite{Chachamis:2004ab}. The mean distance in rapidity among emissions in the BFKL ladder, including higher order collinear contributions, has been recently studied using the Monte Carlo event generator {\tt BFKLex} in~\cite{Chachamis:2015ico}. 

The purely perturbative QCD pomeron has the feature of a cut in the complex angular momentum plane as opposed to a discrete
 pole predicted by the phenomenologically successful Regge theory. As long ago as 1986, Lipatov \cite{lipatov86} pointed out
that the cut can be converted into a series of discrete poles if the running of the QCD  coupling is taken into account and
 that a phase-fixing condition in the infrared region of transverse momentum arising from the non-perturbative
properties of QCD is imposed. This scenario has been studied extensively in ref.\cite{KLR}.

In this letter we combine these two approaches and show that there is a very significant attenuation of the growth of
the BFKL amplitude with rapidity if the rapidity veto is imposed.

\section{Discrete Pomeron in Leading Order}

\begin{figure}
\includegraphics[width=8.0cm]{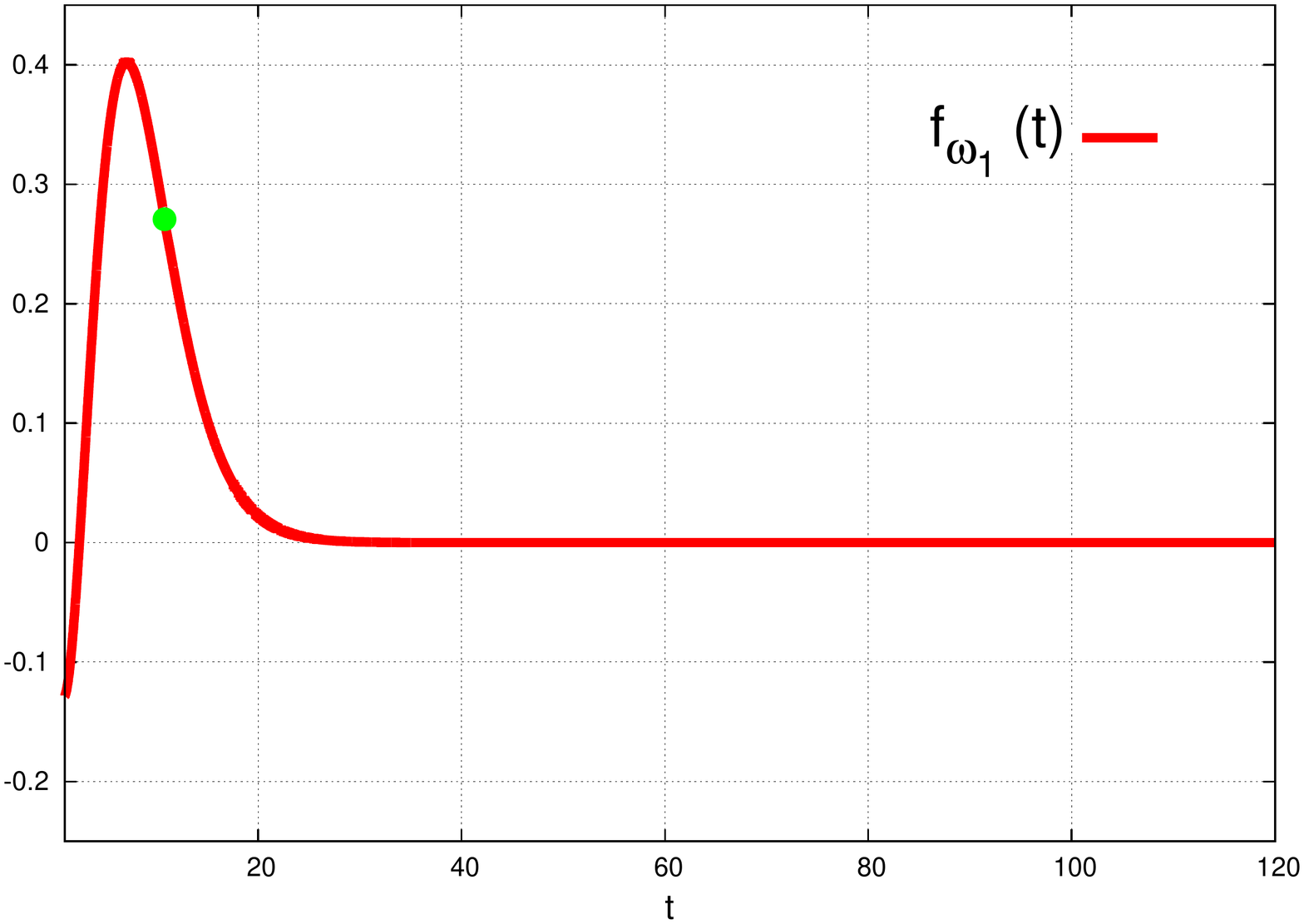}  \includegraphics[width=8.0cm]{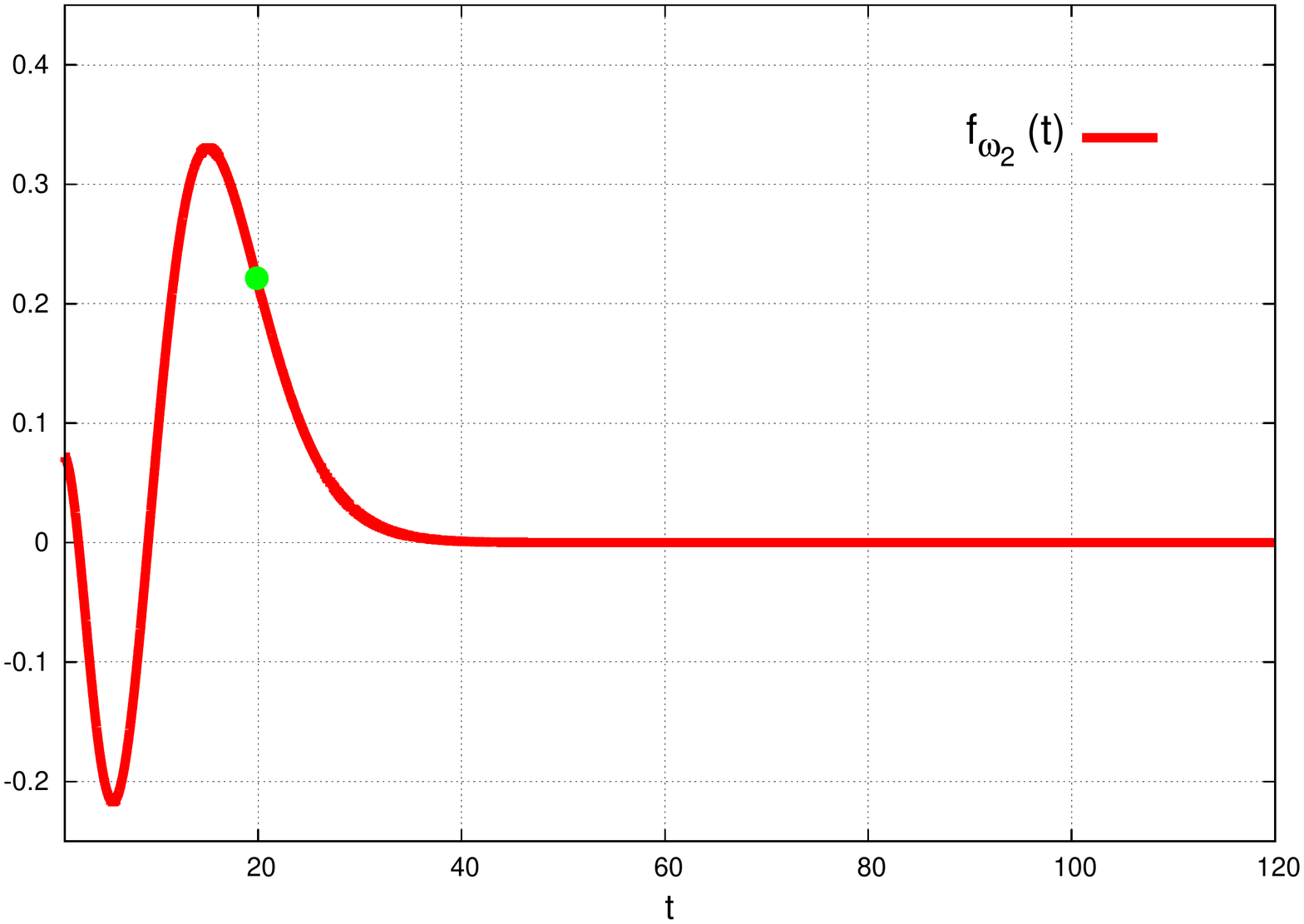} \\
\includegraphics[width=8.0cm]{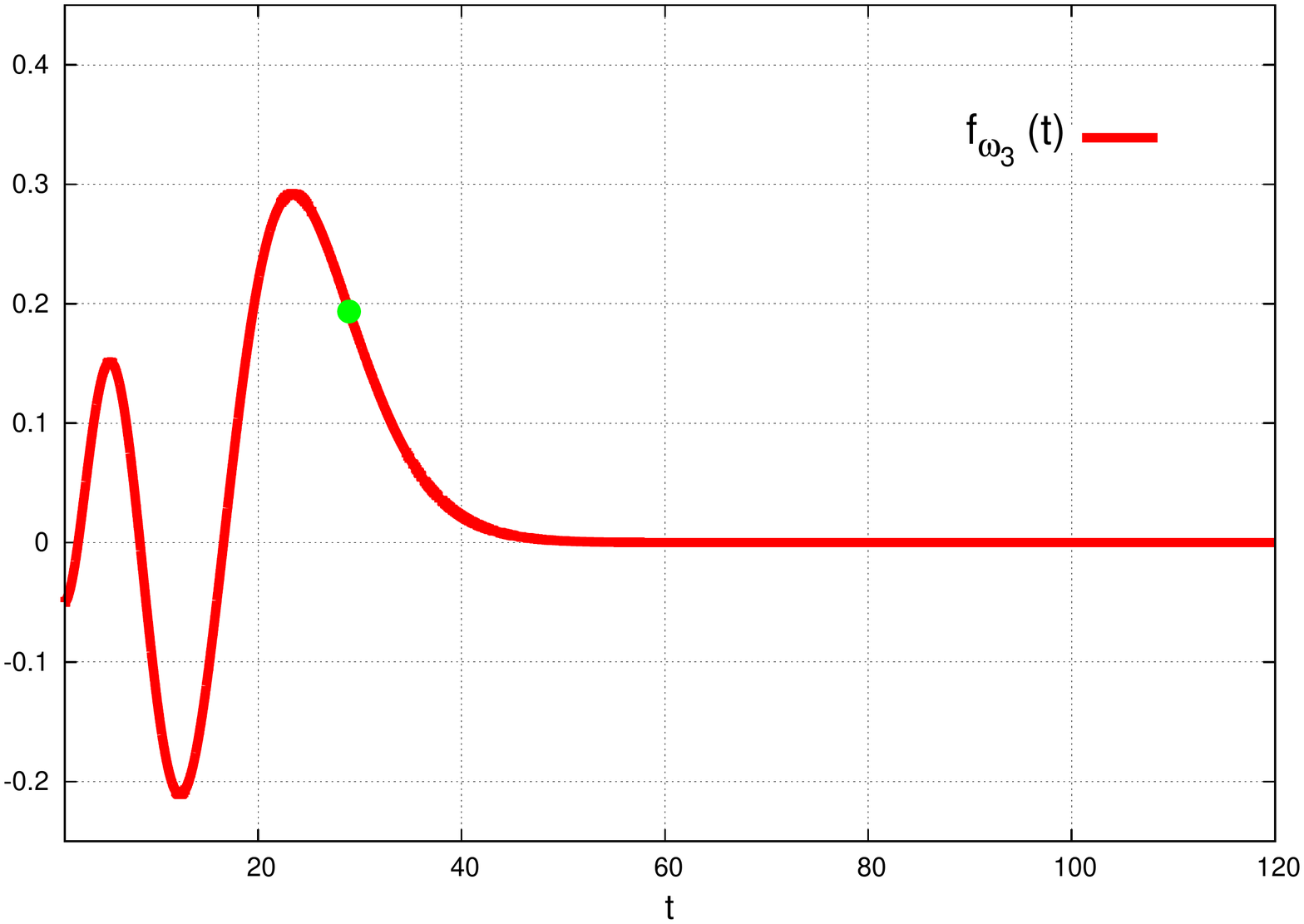} 
\includegraphics[width=8.0cm]{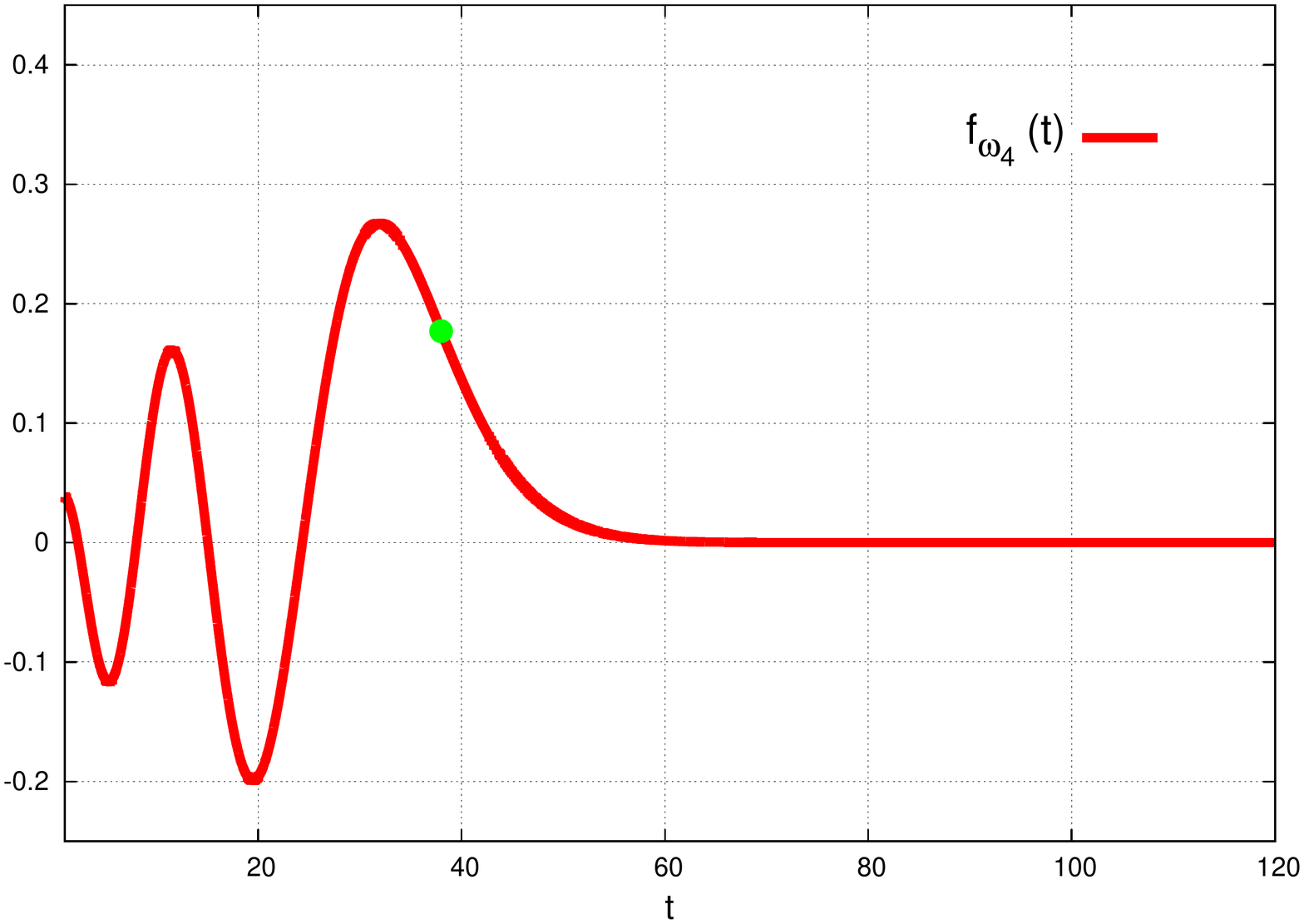} \\ \includegraphics[width=8.0cm]{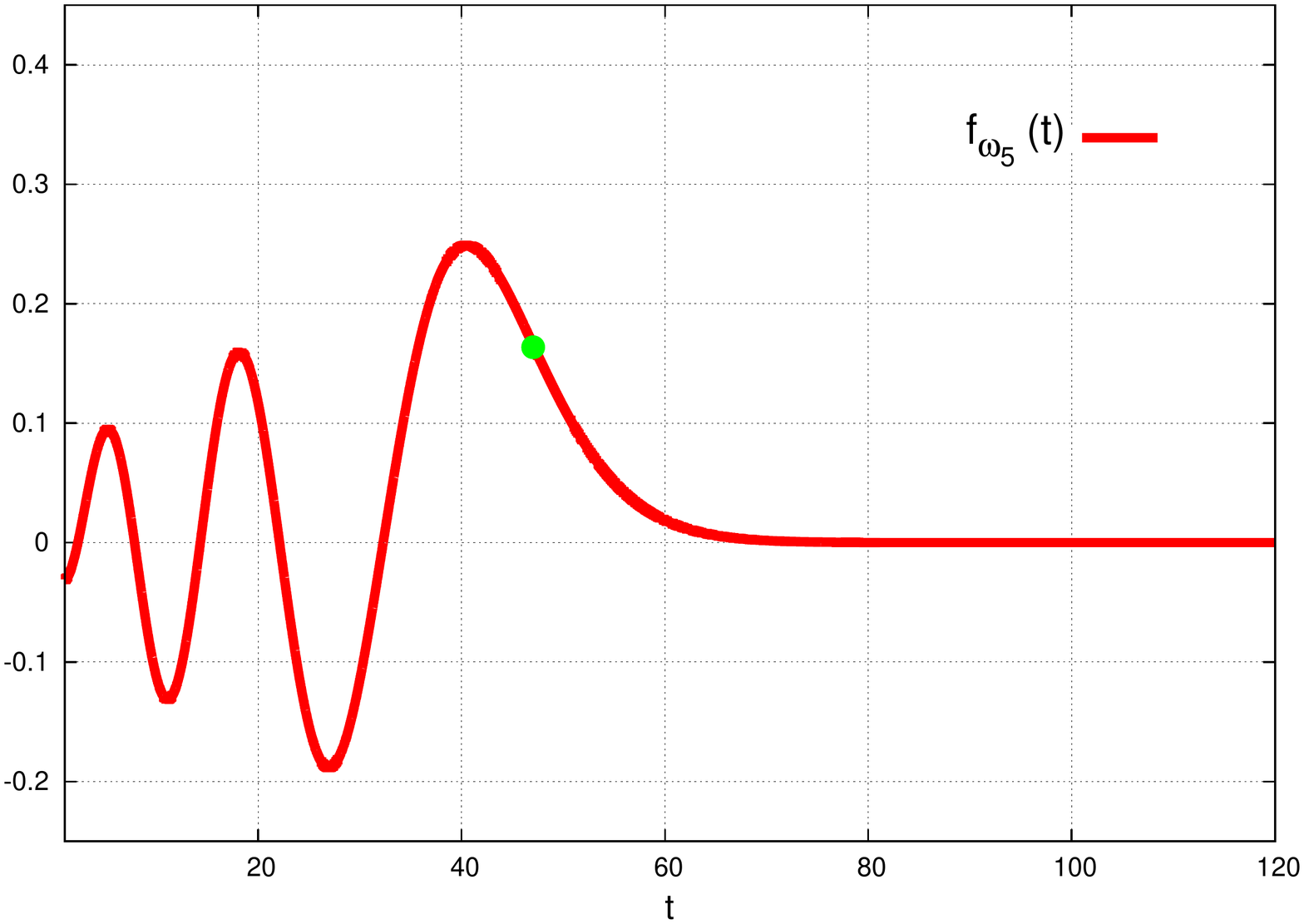} 
\includegraphics[width=8.0cm]{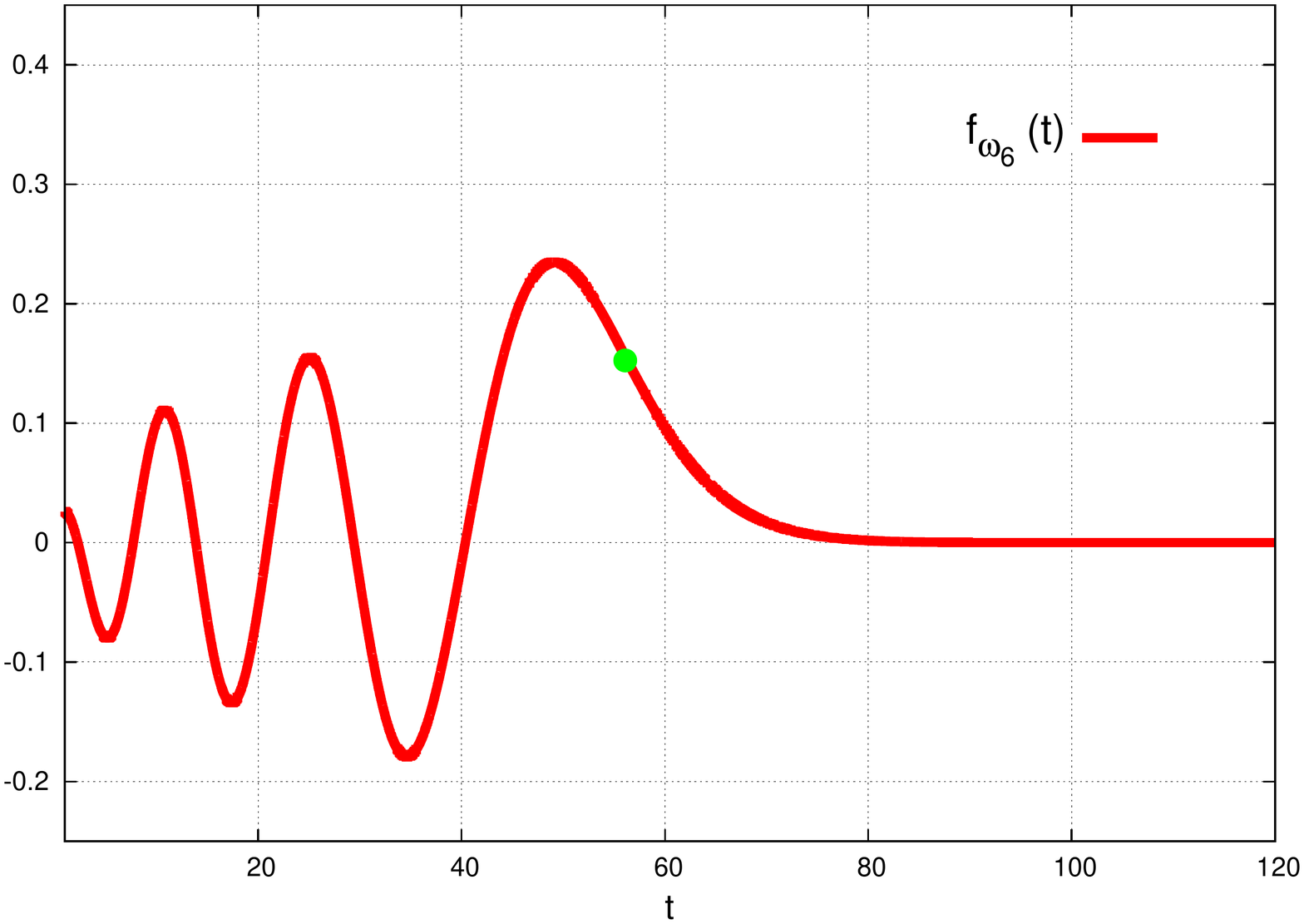}   
\caption{The first six discrete eigenfunctions of the BFKL kernel with running coupling. In each case the green dot
 indicates the value of $t_c \equiv 4\ln2/\betabar\omega$ which delineates between the oscillatory and evanescent
 parts of the eigenfunctions. }
\label{fig1}
\end{figure}

We first reproduce the results for the discrete BFKL pomeron in leading order (LO). For simplicity we neglect the effects of any thresholds arising from massive particles in the running of the coupling and write the running coupling as
\beq \alphabar(t) \ \equiv \ \frac{C_A}{\pi} \alpha_s(t) \ = \ \frac{1}{\betabar t}. \eeq
The Green function, ${\cal G}(Y,t,t^\prime)$, then obeys the equation
\beq \frac{\partial}{\partial Y} {\cal G}(Y,t,t^\prime) \ = \ \int dt^{\prime\prime} \frac{1}{\sqrt{\betabar t}}
 {\cal K}(t,t^{\prime\prime}) \frac{1}{\sqrt{\betabar t^{\prime\prime}}} {\cal G}(Y,t^{\prime\prime},t^\prime), \label{green1} \eeq
where we have introduced the running coupling in such a way as to ensure that the operator on the RHS of (\ref{green1})
is Hermitian. The kernel ${\cal K}$ is the LO BFKL kernel with eigenvalues (in the azimuthally symmetric case) $\chi(\nu)$ where
\begin{eqnarray} 
\chi(\nu)  \ = \ 2 \Psi(1)-\Psi\left(\frac{1}{2}+i\nu\right)-\Psi\left(\frac{1}{2}-i\nu\right).
\end{eqnarray}
In the semi-classical approximation, the normalized eigenfunctions of the kernel with running coupling, with eigenvalue $\omega$,
i.e.
\beq 
 \int dt^{\prime} \frac{1}{\sqrt{\betabar t}}
 {\cal K}(t,t^\prime) \frac{1}{\sqrt{\betabar t^{\prime\prime}}} f_\omega(t^\prime) \ = \ \omega f_\omega(t) \eeq
 are given by
\beq f_\omega(t) \ = \ \frac{|z_\omega(t)|^{1/4}}{\sqrt{\alphabar(t) \chi^\prime\left(\nu_\omega(t) \right)}}
 Ai\left(z_\omega(t)\right), \eeq
where $Ai(z)$ is the Airy function which is regular as $z\to\infty$, 
\beq
\nu_\omega(t) \ = \ \chi^{-1}(\betabar\omega t), 
\eeq  
and 
\beq 
z_\omega(t) \ = \ - \left( \frac{3}{2} \int_t^{4\ln2/\betabar\omega}dt^\prime  \nu_\omega(t^\prime)  \right)^{2/3}.
\eeq
The Airy function is oscillatory for negative argument and the imposition of a fixed phase for such oscillations at some
small value of $t$ leads to a set of discrete eigenfunctions $f_{\omega_n}(t)$. The Green function is then given by
 \beq {\cal G}(Y,t,t^\prime) \ = \ \sum_n f_{\omega_n}(t) f_{\omega_n}^*(t^\prime) e^{\omega_n Y}. \eeq
The first six such eigenfunctions are shown in Fig.\ref{fig1} in the case where an infrared phase of
 $\pi/4$ is assumed at $t=1$. As expected, the $n^{th}$ eigenfunction has $n$ turning points in the oscillatory region.
\section{Introducing a Rapidity Gap Veto}
The imposition of a rapidity gap veto in the kernel follows very much along the lines described in \cite{Schmidt}.
We start by defining the Mellin transform of the Green function with the rapidity $Y$ shifted by $b$, i.e.
 \beq {\cal G}_\omega(t,t^\prime) \ \equiv \int_0^\infty dY e^{-\omega Y} {\cal G}\left((Y+b),t,t^\prime \right) \eeq
which obeys the modified Green function equation
 \beq \omega  {\cal G}_\omega(t,t^\prime) \ = \ \delta(t-t^\prime)+ e^{-b \, \omega}
 \int dt^{\prime\prime} \frac{1}{\sqrt{\betabar t}}
 {\cal K}(t,t^{\prime\prime}) \frac{1}{\sqrt{\betabar t^{\prime\prime}}} {\cal G}(Y,t^{\prime\prime},t^\prime) \eeq
In terms of the discrete eigenfunctions with eigenvalues $\omega_n$, this Mellin transform is given by
 \beq  {\cal G}_\omega(t,t^\prime) \ = \ \sum_n \frac{f_{\omega_n}(t) f^*_{\omega_n}(t^\prime)}{\omega-e^{-b \, \omega} \omega_n } \eeq
and inverting it and shifting the argument of the Green function back to $Y$ we have
\beq {\cal G}(Y,t,t^\prime) \ = \ \int_{\cal C} \frac{d\omega}{2\pi i} e^{\omega(Y-b)}   \sum_n \frac{f_{\omega_n}(t) f^*_{\omega_n}(t^\prime)}{\omega-e^{-b \, \omega} \omega_n }. \label{invmell} \eeq
The term $\left(\omega-e^{-b \, \omega} \omega_n\right)^{-1}$ 
has a pole  at \beq \omega \ =  \ \frac{W(b \, \omega_n)}{b}  \ \equiv \ \overline{\omega}_n \eeq
where $W(x)$ is the Lambert $W$-function (defined as the solution to $ x \ = \ W(x)e^{W(x)}$), with 
residue $\left(1+b \, \overline{\omega}_n\right)^{-1} $ so that finally our expression for the Green function  with rapidity gap veto $b$ is given by
\beq 
{\cal G}(Y,t,t^\prime) \ = \    \sum_n  e^{\overline{\omega}_n(Y-b)}
\frac{f_{\omega_n}(t) f^*_{\omega_n}(t^\prime)}{1+b \, \overline{\omega}_n }. \label{green2} 
\eeq

It is interesting to note that we may re-express this more simply as
\beq 
{\cal G}(Y,t,t^\prime) \ = \    \sum_n  e^{\overline{\omega}_nY}
\overline{f}_{\overline{\omega}_n}(t) \overline{f}^*_{\overline{\omega}_n}(t^\prime) \label{green3} 
\eeq
where $\overline{f}_{\overline{\omega}}(t)$  are the eigenfunctions of the kernel with running coupling
normalized as
 \beq \int dt \overline{f}_{\overline{\omega}}(t)\overline{f}^{\, *}_{\overline{\omega^\prime}}(t) \ = \ \delta\left(\overline{\omega}-\overline{\omega^\prime} \right) \eeq
so that in order to account for this normalization we have
\beq f_\omega(t) \ = \ \eta(\overline{\omega}) \overline{f}_{\overline{\omega}}(t) \eeq
where
 \beq \left| \eta(\overline{\omega})\right|^2 \ = \ \frac{d\overline{\omega}}{d\omega} \ = \ 
\frac{e^{-b \, \overline{\omega}}}{1+b \, \overline{\omega}},\eeq
which matches the factor 
\beq
\frac{e^{- b \bar{\omega}_n}}{1+ b \, \bar{\omega}_n}
\eeq
in each term in the sum on the RHS of eq.~(\ref{green2}).

\section{Results}

\begin{figure}
\centerline{\includegraphics[width=15.0cm]{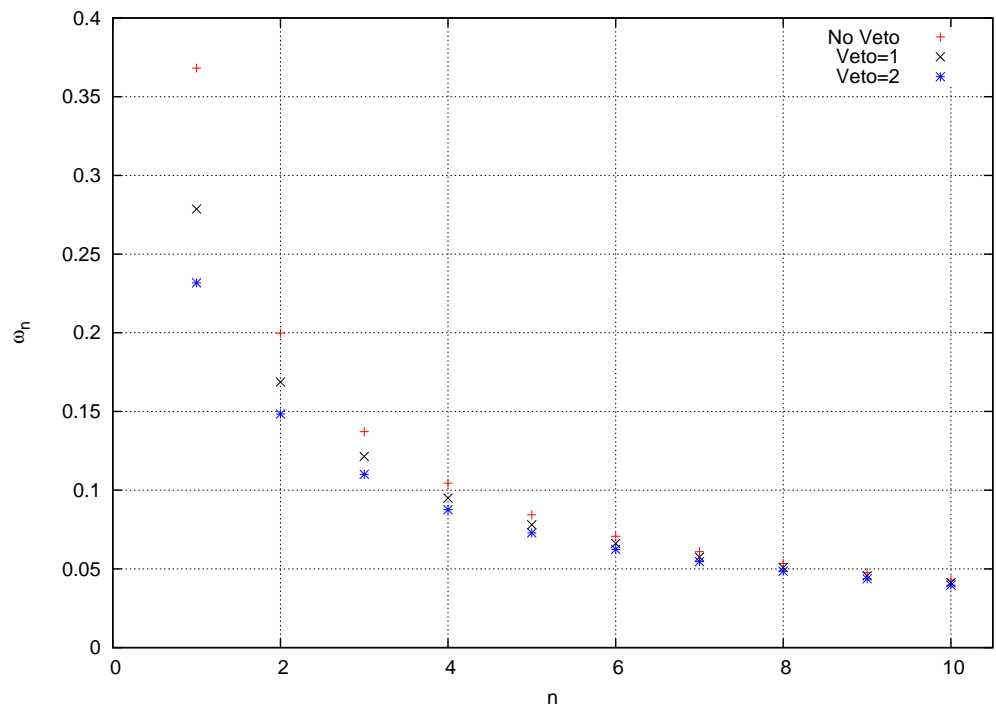}} 
\caption{The  reduction of the first 13 effective eigenvalues from the imposition of a rapidity veto $b=1$ (black crosses)
 and  $b=2$ (blue stars). }
\label{fig2}
\end{figure}

We can see from eq.~(\ref{green2}) that the imposition of a rapidity gap veto, $b$,  attenuates the Green function in two different ways. The first is the simple shift of $Y$  to $Y-b$. The second is the replacement of the eigenvalues $\omega_n$
 by their reduced values $\overline{\omega}_n$. This reduction is shown in Fig.~\ref{fig2} and we notice that the effect
is much larger for the leading eigenvalues than the subleading. This immediately tells us that the effect of the rapidity gap veto
is largest for very large values of $Y$ for which we expect the Green function   to be dominated by the leading eigenvalue.
 On the other hand, it has been shown in \cite{KLRlatest} that when $t$ increases beyond $t_c=4\ln2/\betabar\omega_1$
 the residue of this leading pole becomes evanescent and eventually the first subleading pole becomes dominant. We therefore
conclude that the effect on the Green function is reduced as the values of $t$ and $t^\prime$ increase. This effect, however, is very
slow and one has to consider very substantial values of $t$ before such behaviour becomes manifest.

\begin{figure}
\centerline{\includegraphics[width=15.0cm]{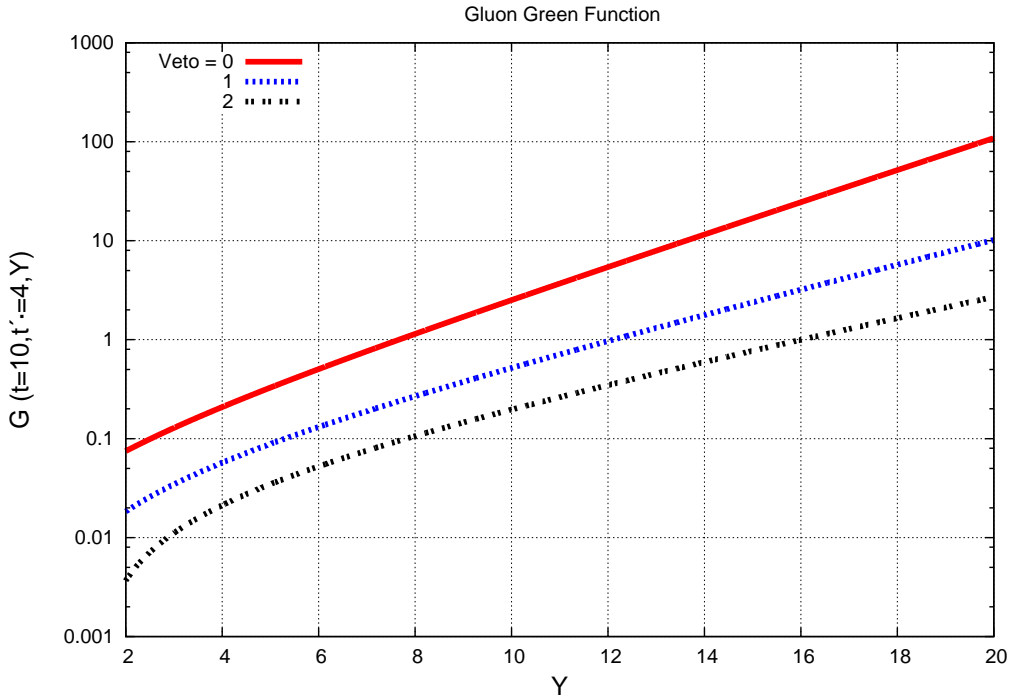}} 
\caption{The rapidity dependence of the Green function without a rapidity veto (red solid), with rapidity gap veto 1 (blue dotted) and rapidity veto 2 (black dotted). }
\label{GGFvsY}
\end{figure}

For a typical pair of values of $t$ and $t'$, namely $t=10$, $t'=4$, we show in Fig.~\ref{GGFvsY} the 
growth in the Green function with $Y$ for the case of rapidity gap veto 0, 1 and 2 and note that the 
imposition of a veto of two units, preferred in~\cite{FRS}, can reduce the value of the Green function by an order of magnitude. Whereas the overall value can often be absorbed into a redefinition of the impact factors of the scattering particles, we note that the divergence of these lines indicates that the growth of diffractive cross-sections with increasing rapidity gap is noticeably reduced when such a veto is imposed. 

\begin{figure}
\centerline{\includegraphics[width=15.0cm]{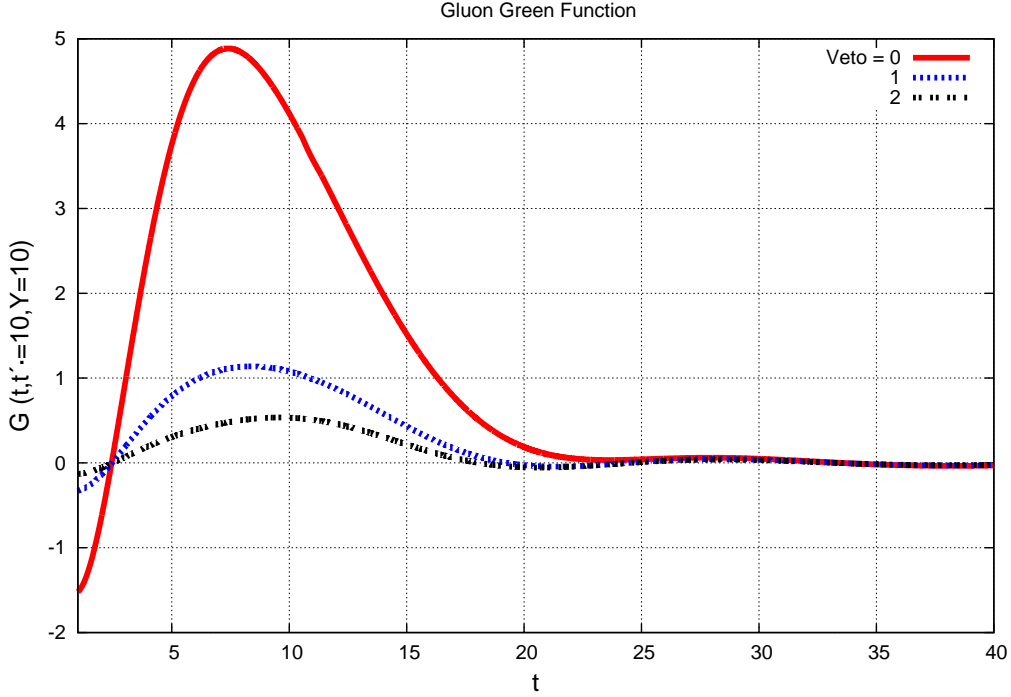}} 
\caption{The transverse momentum $(t)$ dependence of the Green function without a rapidity veto (red solid), with rapidity gap veto 1 (blue dotted) and rapidity veto 2 (black dotted) for $t'=10$ and $Y=10$. }
\label{GGFvstY10}
\end{figure}

Finally, in Fig.~\ref{GGFvstY10} we examine the effect of the rapidity veto on the $t$ dependence of the Green function for $Y=10$. We note that the distinct peak in the vicinity of $t=t'$ which is present in the 
case of zero veto is substantially suppressed and broadened when a rapidity-gap veto is introduced. For 
$Y=0$ we get a $\delta$-function which can be seen from the completeness relation of the set of eigenfunctions $\bar{f}$. As $Y$ increases this $\delta$-function is broadened, giving rise to the distribution in transverse momentum which is broader in the center than at the ends (the so-called ``Bartels cigar"~\cite{BartelsCigar}). Fig.~\ref{GGFvstY10} shows that with the imposition fo a rapidity-gap veto we expect this ``cigar" to become fatter. 
  
\section{Conclusions} 

Higher order corrections to the BFKL equation are very important for theoretical and phenomenological 
studies of QCD at high energies. It is well-known that the largest portion of the next-to-leading corrections are due to running of the coupling effects and collinear contributions. Both have been treated in the present work using the discrete 
pomeron approach together with the introduction of a veto in the relative rapidities of the emitted gluons in the BFKL gluon Green function. The rapidity veto samples the region of phase space corresponding to collinear emissions already at a value of two units of rapidity. We have shown how to implement this veto when infrared boundary conditions are imposed with a running coupling such that the singularities in the complex angular momentum plane are only Regge poles and no branch cuts. This is a novel approach which should be most relevant when investigating observables characterized by external scales which are not too hard.  It will be interesting to put these ideas to work at different observables in hadrons collisions such as those being tested at the Large Hadron Collider.

\section*{Acknowledgements}
One of us (DAR) wishes to thank the Institute for Theoretical Physics
at the Autonomous University of Madrid for its hospitality during the time
that this work was carried out,  as well as the Leverhulme Trust for an Emeritus Fellowship.  ASV acknowledges support from the Spanish Government (MICINN (FPA2015-65480-P)) and from the Spanish MINECO Centro de Excelencia Severo Ochoa Programme (SEV-2012-0249). We would like to thank the university of Alcal{\'a} de Henares for the use of their facilities at the Colegio de M{\'a}laga.

\end{document}